\documentclass[twocolumn,prl,aps,floatfix]{revtex4}
\usepackage{graphicx} %figuras
\usepackage{color}
\begin{document}
%\preprint{}
\title{ Growth laws and self-similar growth
% scaling
 regimes of coarsening two-dimensional foams:\\
    Transition from dry to wet limits} %Title of paper
\author{Ismael Fortuna}
\email[]{ismaelfortuna@gmail.com}
%\thanks{}
\affiliation{Instituto de  F\'{\i}sica, Universidade Federal do Rio Grande do Sul\\ C.P. 15051, 91501-970 Porto Alegre, RS Brazil}
\author{Gilberto L. Thomas}
%\email[]{glt@if.ufrgs.br}
%\thanks{}
\affiliation{Instituto de F\'{\i}sica, Universidade Federal do Rio Grande do Sul\\ C.P. 15051, 91501-970 Porto Alegre, RS Brazil}
\author{Rita M.C. de Almeida}
\email[]{rita@if.ufrgs.br}
%\thanks{}
\altaffiliation{Instituto Nacional de Ci\^encia e Tecnologia: Sistemas Complexos}
\affiliation{Instituto de F\'{\i}sica, Universidade Federal do Rio Grande do Sul\\ C.P. 15051, 91501-970 Porto Alegre, RS Brazil}
\author{Fran\c{c}ois Graner}
%\email[]{francois.graner@curie.fr}
%\thanks{}
%\altaffiliation{Present address: Mati\`ere et Syst\`emes Complexes, Universit\'e Paris Diderot, CNRS UMR 7057, 10 rue Alice Domon et L\'eonie Duquet, 75205 Paris Cedex 13, France}
\affiliation{
 ``Polarit\'e, Division et Morphogen\`ese",
Laboratoire de G\'en\'etique et Biologie du D\'eveloppement, Institut Curie, 26 rue d'Ulm,
F-75248 Paris Cedex 05, France}
\affiliation{Mati\`ere et Syst\`emes Complexes, Universit\'e Paris Diderot, CNRS UMR 7057, 10 rue Alice Domon et L\'eonie Duquet, F-75205 Paris Cedex 13, France}

\date{\today}

\begin{abstract}
We study the topology and geometry of two dimensional coarsening foams with arbitrary liquid fraction.
To  interpolate between the dry limit described by von Neumann's law, and the wet limit described by Marqusee equation, the relevant bubble characteristics are the Plateau border radius and a new variable,
the effective number of sides.
We propose   an equation for the individual bubble  growth rate  as the weighted sum of the growth through bubble-bubble interfaces and through bubble-Plateau borders interfaces.
The resulting prediction is successfully tested, without adjustable parameter, using extensive bidimensional Potts model simulations.
 Simulations also show that a self-similar growth regime is observed at any liquid fraction and determine how the average size growth exponent, side number distribution and relative size distribution interpolate between the extreme limits.
 Applications include concentrated emulsions, grains in polycrystals  and other domains with coarsening driven by curvature.
\end{abstract}


\maketitle %\maketitle must follow title, authors, abstract and \pacs

% Introduction
Liquid foams, namely gas bubbles separated by a continuous liquid phase,
are ubiquitous \cite{weaire_book,livre_mousse}.
In floating foams as beer heads, ocean froths or pollutant foams, the fraction $\phi$ of their volume occupied by the liquid decreases with height, varying from a dry foam at the top  to a bubbly liquid at the foam-liquid interface.

Since pressure can differ from one bubble to another, gas slowly diffuses. Some bubbles disappear and, as no new one is created, the average size increases.
Foam coarsening is analogous to that of concentrated emulsions, grains in polycrystals, or two-phase domains where interface dynamics is driven by curvature.
Its dynamics depends mainly on $\phi$, up to a material-specific time scale determined by the foam physico-chemistry  \cite{weaire_book,livre_mousse}.

Understanding foam coarsening requires two different levels. First, the {\it individual bubble growth law}, which rules a bubble's growth rate according to its size or shape. This law can be stated as a static geometry problem, and may be obtained analytically or  by detailed bubble shape simulation. Second,  the effect of such individual growth on the {\it statistics of the foam}, \emph{i.e.}, bubble size and topology distributions, requires statistical theories or large bubble number simulations.

In the {\it very dry} limit $\phi \to 0$, bubbles are polyhedra with thin curved faces meeting by three along thin lines called Plateau borders. Coarsening in that limit has been  investigated  experimentally, numerically and theoretically in two (2D) \cite{Ne52,Stavans89,St93,pignol,Iglesias91} and later in three dimensions (3D) \cite{WEO00,KC02,Streitenberger,hilgenfeldt04,TAG06,macpherson07,LMCCGGD10}.

 In the {\it very wet} limit $\phi \to 1$, bubbles are round, dispersed in the liquid and far from each other, forming a ``bubbly liquid" rather than a foam \emph{stricto sensu}. Their  coarsening follows ``Ostwald-Lifschitz-Slyozov-Wagner" ripening,  in 3D \cite{Os1902,LS58} and later in 2D \cite{Marq84,YEGG92}.

In both limits, the foam eventually reaches a self-similar growth regime: statistical distributions of face numbers and relative sizes become invariant. Only the average size $\langle R \rangle$ grows in time, as a power law $\langle R \rangle \sim t^\beta$, with $\beta=1/2$ in the dry limit and $\beta=1/3$ in the wet one, reflecting that the underlying physical processes are different. The number $N$ of bubbles  thus decreases as $t^{-2\beta}$ in 2D and $t^{-3\beta}$ in 3D.
  The growth law for intermediary liquid fractions has been addressed in experiments \cite{LCDMCGG07} and simulations  \cite{BW92,HWB95}, but still lacks a unified theoretical description.

 Here we address the 2D case. We propose a growth law to interpolate for $0<\phi<1$, with two parameters (diffusion coefficients) which are determined in each limit. To test our prediction on 2D foam coarsening experiments is difficult, because we are not aware of any study where $\phi$ is systematically varied and precisely measured, or even rigorously defined.
We rather use  numerical simulations based on Potts model,  suitable for  large bubble numbers  \cite{zollner,AGS89,GAG90,TAG06}. Beside testing our prediction, simulations also show that,  for any $\phi$, side number and relative size distributions  reach a self-similar growth regime
% steady states
 where $\langle R \rangle$ grows as $ t^\beta$. Values of $\beta$ interpolate between $1/2$ and $1/3$.

 % Theory
In the 2D dry limit, gas diffuses through neighbor bubbles walls, due to the pressure difference between bubbles related with wall curvature. The walls are curved because they   meet at threefold vertices,  forming equal angles of $2\pi/3$.
Each vertex is responsible for a turn of $\pi/3$ in the vector tangent to the bubble perimeter.
Consequently bubbles with $5$ sides or less are convex, while bubbles  with $7$ sides or more are concave, so that the   walls curvature  plus a turn of $\pi/3$ at each vertex sums up to $2\pi$ \cite{livre_mousse}. The resulting growth dynamics is von Neumann's law \cite{Ne52}:
\begin{equation}
\label{vonneumann}
\frac{\mbox{d}a^{i}}{\mbox{d}t}=-D_{d} \left(2\pi -n^{i}\frac{\pi}{3}\right) =\frac{\pi D_{d}}{3} \left(n^{i}-6\right) \ ,
\end{equation}
where $a^{i}$ and $n^{i}$ are, respectively, area and  number of sides of the $i^{th}$ bubble; $t$ is time; $D_{d}$ depends on the foam composition and is expressed in m$ ^2$s$^{-1}$ as a diffusion coefficient. Remarkably, the rhs of eq. (\ref{vonneumann}) involves only the bubble's number of sides and not its size or shape. At any time, bubbles with $n^{i}<6$ shrink while bubbles with  $n^{i}>6$ grow.
Since for topological reasons  the average bubble number of sides is 6 \cite{graustein, livre_mousse}, eq. (\ref{vonneumann}) is compatible with  gas volume conservation in the whole foam.

In the 2D wet limit, gas bubbles are dispersed in a liquid matrix. Dynamics is a consequence of pressure difference in the gas contained in a bubble or dissolved in the liquid.  This pressure difference   is proportional to the wall curvature, which for a circular bubble is  the inverse of its radius, $R^i$.
Marqusee \cite{Marq84} wrote the growth law  using only $R^i$ :
\begin{equation}
\label{marqusee1}
\frac{\mbox{d}a^{i}}{\mbox{d}t}=2\pi R^i\frac{\mbox{d}R^{i}}{\mbox{d}t}=2\pi D_w
f(R^i),
\end{equation}
where$D_w$ is another diffusion coefficient-like constant as above, and
\begin{equation}
\label{marqusee2}
f(R^i) =  \frac{R^i}{\xi} \frac{K_1\left(\frac{R^i}{\xi}\right)}{K_0\left(\frac{R^i}{\xi}\right)}\left[\frac{1}{R_c}-\frac{1}{R^i}\right] \ ,
\end{equation}
where $K_j$s are $j^{\mbox{th}}$ order modified  Bessel functions of second kind; $\xi$ is the screening length (roughly, the typical distance beyond which bubbles do not feel the influence of each other); and $R_c$ is the critical radius for which there is no growth, calculated by imposing  total gas volume conservation, \emph{i.e.} $\frac{\mbox{d}}{\mbox{d}t}\sum_{i}a^{i}=0$.
At any given time, bubbles with radius smaller than $R_c$  lose gas while those with radius larger than $R_c$ gain gas.

\begin{figure}
       \includegraphics[angle=0.0,width=0.6\linewidth]{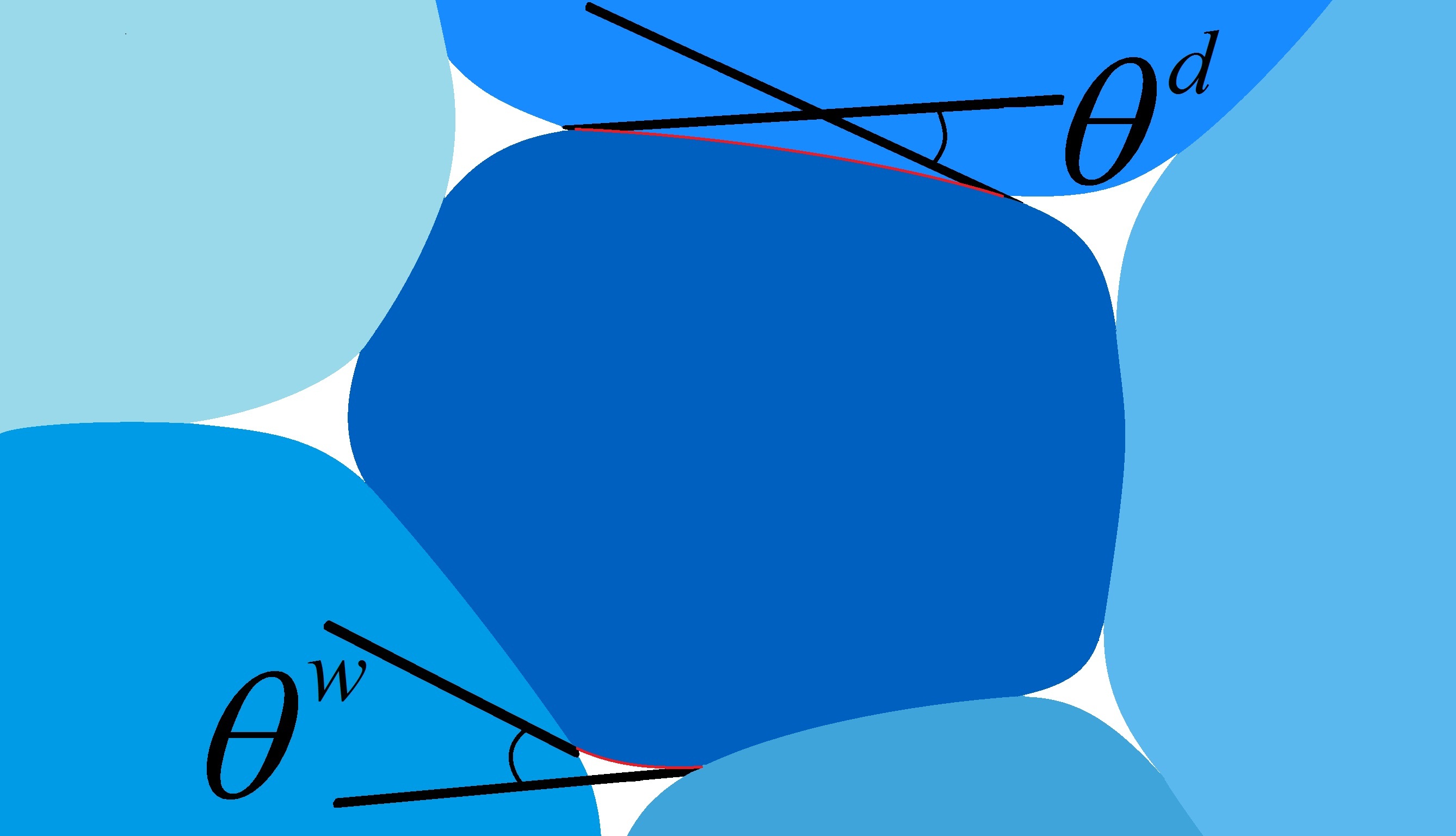}
\caption{(Color online) Angles  $\theta_{d}$ or $\theta_{w}$  measure the change in direction carried by a dry or wet interface.  Here $n_{eff}$ is slightly larger than $5$, the bubble is rather dry, $\Theta_{d} > \Theta_{w}$, $R_w < R$. Each white region represents a Plateau border and is simulated as several liquid drops (not drawn here).}
\label{angles}
\end{figure}

Interpolating between eqs. (\ref{vonneumann}) and (\ref{marqusee1}) seems difficult because they use very different variables: number of sides $n^{i}$ and radius $R^{i}$.
However, both equations involve the product of curvature times length ({\it i.e.} angle, Fig. \ref{angles}) of the interfaces through which the gas diffuses.
We propose  that for any  $\phi$ the bubble growth rate is simply the superposition of growth through the  interfaces shared  either directly with other bubbles or with Plateau borders. It can thus be calculated as  the weighted average of eqs. (\ref{vonneumann}) and (\ref{marqusee1}). The weights are  fractions of $2\pi$ angle carried by  dry or wet parts of the $i^{th}$ bubble perimeter, $\Theta_{d}^{i}$ or $\Theta_{w}^{i}$, which  are  sums of  angles $\theta_{d}$ or $\theta_{w}$ carried, respectively, by all dry or wet interfaces of the bubble
 (Fig. \ref{angles}), such that $\Theta_{d}^{i} + \Theta_{w}^{i}=2\pi$. We now detail how to perform this linear superposition, which turns out to be unexpectedly successful at all $\phi$s.

In dry foams, at each vertex $\theta_{d}=\pi/3$, so that $\Theta_{d}^{i}=n^{i}\pi/3$.
 In wet foams, $\Theta_{w}=2\pi$. To interpolate, we characterize the bubble $i$  by introducing its {\it effective  number of sides}  defined for any $\phi$ as:
\begin{equation}
\label{n_eff}
n_{eff}^{i} = \frac{3}{\pi} \Theta_{w}^{i} = 6 -  \frac{3}{\pi}  \Theta_{d}^{i} \ .
\end{equation}
Although $n_{eff}^{i} $ conveys no more information than $\Theta_{d}$ or $\Theta_{w}$, its value is intuitive: $n_{eff}^{i}=n^{i}$ for a polygonal dry bubble; $n_{eff}^{i}=6$ for a circular wet bubble;  $1-n_{eff}^{i}/6$ or $n_{eff}^{i}/6$ are exactly  the fractions of $2\pi$ carried by either dry or wet interfaces, respectively.
Similarly,  we characterize the bubble $i$ by the curvature radius $R_{w}^{i}$ of its Plateau borders;  $R_w^{i}=0$  for a dry bubble and  $R_w^{i}=R^{i}$ for a wet bubble. With these variables $n_{eff}^{i}$, $R_w^{i}$ we can weight eqs. (\ref{vonneumann}) and (\ref{marqusee1}) and interpolate for any $\phi$:
 \begin{equation}
\label{growth1}
\frac{\mbox{d}a^{i}}{\mbox{d}t}=
\frac{\pi}{3} \;
\left[ D_{d} (n_{eff}^i-6) + D_{w} n_{eff}^i f(R_{w}^{i}) \right] \ .
\end{equation}

%Methods
To test eq. (\ref{growth1}) we have implemented extensive Potts model  simulations in the spirit of  \cite{TAG06}.
Like in experimental pictures, a simulation represents gas bubbles and liquid phase as connected regions on a square lattice of pixels.
To ensure that liquid is evenly distributed between  Plateau borders  (fig. \ref{angles}), and that  $\phi$ is conserved, we represent the liquid phase as a fixed  number of   tiny ``liquid drops" of fixed area,
and each Plateau border contains several of these drops.
Each interface between gas bubbles represents a thin film made of
two gas-liquid interfaces which mildly repel each other. We then assign to the interfacial energy between gas bubbles a value $(2-\varepsilon)$ times  the interfacial energy between gas bubble and liquid drop. Experimentally, in soap foams $\varepsilon$ is typically of order $10^{-3}$. Its precise value is not crucial in what follows, as long as $\varepsilon \ll 1$ (and  $\varepsilon >0$ to avoid numerical unstabilities).
 Here we choose   $\varepsilon=0.01$ as a compromise between realism and computing speed.
 Since liquid drops are free to cluster together, we assign to drop-drop interfaces an energy $10^3$ times smaller. The liquid is sucked in the Plateau borders
(for reviews see  \cite{livre_mousse,Ber99}).

There are $\cal{N}=$ $8944^2$ sites; each site $s$  is assigned with a label, $S(s)$.
There are $N$ bubbles ($S=1$ to $N$) and $N_w$ drops ($S=N+1$ to $N+N_w$), assigned with a type  $\tau=1$ or 0, respectively.
A configuration has an energy:
\begin{equation}
\label{energy}
E\! = \! \sum_{s=1}^{\cal{N}} \! \sum_{s'=1}^{36}  J\left(\tau,\tau' \right)[1 - \delta_{S,S'}] +  \lambda \!\!\!  \sum_{S=N+1}^{N+N_w} \left[a(S)-a_{t}\right]^2,
\end{equation}
where $s'$ stands for the sum over the first $36$ neighbors of the site $s$, to avoid pinning to the grid \cite{HGSG91} and to extend the range of the disjoining pressure up to three pixels; $S$ and $S'$ are the labels of sites $s$ and $s'$, respectively; $\delta$ is the Kronecker symbol; $\tau$ and $\tau'$ are the types of $S$ and $S'$, respectively; $J(1,1)=1.99\;  J(1,0)$, $J(1,0)=J(0,1)=0.7$, and $J(0,0)=0.001$ are the interfacial energies;
 $a(S)$ is the current area of liquid drop $S$; $a_{t}=8$ pixels is a target area common to all drops; $\lambda= 9$ penalises any deviation from $a_{t}$.

Simulations begin with $N_0=2 \times 10^5$ gas bubbles randomly dispersed over the grid, with smooth interfaces and a normal distribution of areas around the average $\frac{\cal{N}}{N_0}$. For $\phi>0$, in the initial configuration each vertex contains at least one drop. The total number of drops and the initial average area of gas bubbles are set accordingly to the desired $\phi$.

The simulation dynamics follows Monte Carlo method. We randomly choose a site, temporarily change its label to the value of one of its neighbors, and calculate the change in energy $\Delta E$. If $\Delta E\leq 0$ this relabeling is accepted. If $\Delta E>0$ the change is accepted with probability $\mbox{exp}(-\Delta E/T)$, where $T$ is the fluctuation allowance, here taken as $T=3$ to escape possible metastable states.

 We measure diffusion coefficients as follows. We first perform a simulation in the dry limit $\phi=0$. Plotting the rhs and lhs of eq.  (\ref{vonneumann}) determines by linear regression $D_d=1.68$. We then perform a simulation in the wet limit $\phi=0.9$. Plotting the rhs and lhs of eq.  (\ref{marqusee1}) determines by linear regression $D_w=0.78$.
  With these parameters, together with area and wet and dry perimeters for each bubble from simulations, eq. (\ref{growth1}) predicts without any adjustable parameter  each bubble growth  for any intermediary $\phi$.

 We measure angles on a square grid as follows. The simulation is halted.
We add over all \textit{dry} interfacial pixels of   bubble $i$ the probability that it would grow (or shrink) over (yielding to) other bubbles, calculated by the Monte Carlo method.
This determines  the growth rate of bubble $i$ through its dry interfaces, $G_{d}^{i}$, and thus
$\Theta_{d}^{i} = D_{d} G_{d}^{i}$. We then obtain $ \Theta_{w}^{i} = 2\pi - \Theta_{d}^{i}$,
and  $R_{w}^{i}=\frac{P_{w}^{i}}{\Theta_{w}^{i}}$ where $P_{w}^{i}$ is the wet interface length.
During this measure,  no change is performed. The simulation then resumes.

\begin{figure}
       \includegraphics[angle=0.0,width=1.0\linewidth]{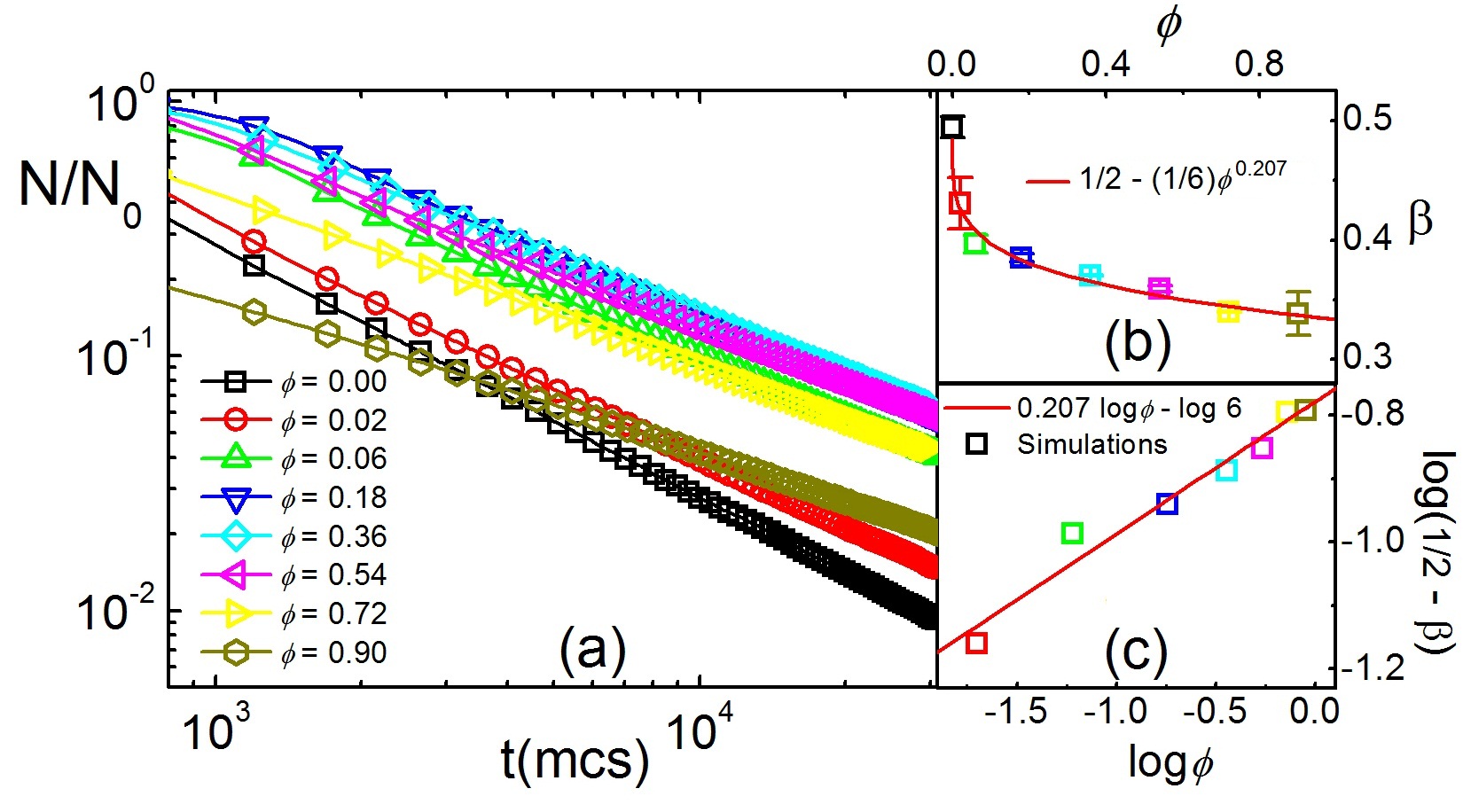}
\caption{(Color online) (a) Evolution of the number of gas bubbles for different values of  $\phi$, in log-log plot. (b,c) Power law exponent  $\beta$  {\it versus} $\phi$, in (b) linear and (c) log-log scales. The red line is $\beta = 1/2-\phi^{0.207}/6$.}
\label{Nvst}
\end{figure}

% Results
We run simulations with $\phi =$ $0$, $0.02$, $0.06$, $0.18$, $0.36$, $0.54$, $0.72$ and $0.90$.
For any $\phi$, the evolution of the   gas bubble number has a  power law behavior (Fig. \ref{Nvst}(a)), which is  compatible with self-similar growth
% scaling
regimes.
 Fig. \ref{Nvst}(b,c) shows that the power law exponent $\beta$ varies as $\beta(\phi) \approx 1/2-\phi^{0.2}/6$.
Thus $\beta(\phi)$ decreases continuously   from $1/2$ to $1/3$, the expected limit values, with d$\beta/$d$\phi$ diverging at $\phi \to 0$.
 It would be interesting to explain theoretically this variation of $\beta$ with $\phi$.

\begin{figure}
       \includegraphics[angle=0.0,width=0.95\linewidth]{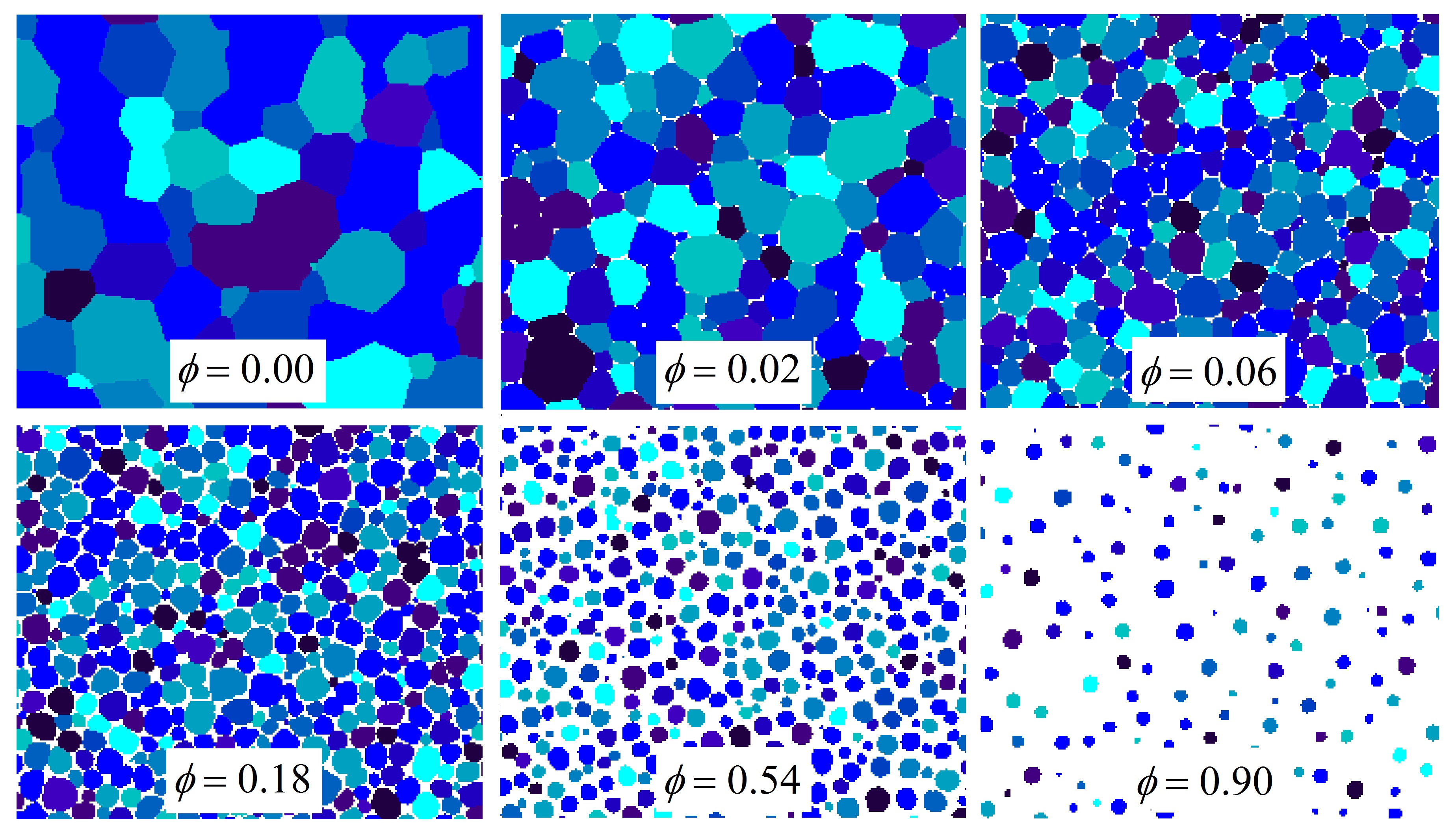}
\caption{(Color online) Snapshots after 20,000 MCS for $\phi=0$, $0.02$, $0.06$, $0.18$, $0.54$, and $0.90$.}
\label{Snapshots}
\end{figure}

Fig. \ref{Snapshots} shows snapshots for different $\phi$s after 20,000 Monte Carlo Steps (MCS). The liquid accumulates at the vertices for small $\phi$s, and, as it increases, liquid goes also between bubbles.
The ratio $d / \xi$, where $d$ the typical distance between bubbles, estimated as
$d = 2  (\sqrt{ {\cal{N}}/{(N\pi)}} - \sqrt{ {\langle a\rangle}/{\pi}})$, and $\xi$  the screening length (eq. \ref{marqusee1}), increases with $\phi$ (Fig. S1 of \cite{suppmat}). At $\phi = 0.54$, $d / \xi > 1$, and at $\phi=0.90$ the bubbles are not touching each other.

\begin{figure}
       \includegraphics[angle=0,width=0.9\linewidth]{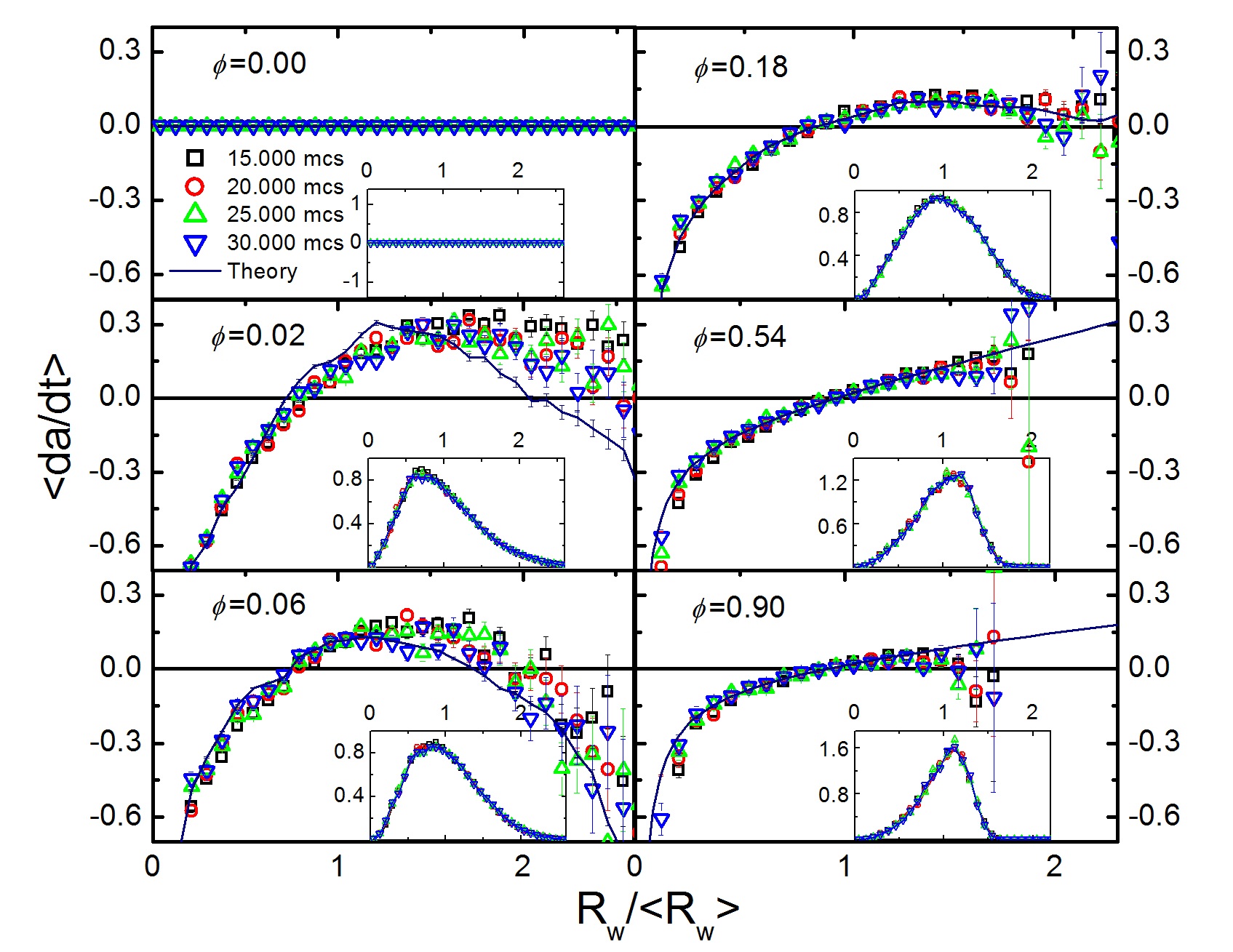}
\caption{(Color online)  Gas bubbles growth rate \emph{versus}   $R_{w}/\langle R_{w}\rangle$ at different times of the self-similar growth
% scaling
 regime  for  $\phi=0.02$ to 0.90. Insets: distribution function of $R_{w}/\langle R_{w}\rangle$.}
\label{Gr_vs_rPB}
\end{figure}

Fig. \ref{Gr_vs_rPB} presents the average area growth rate of gas bubbles \emph{versus} $R_{w}$, and in insets the distribution function  of   $R_{w}/\langle R_{w}\rangle$. The superposition of plots taken at different times indicates  a self-similar growth
% scaling
 regime. The agreement with the theoretical prediction  (eq. \ref{growth1}) is excellent. For $\phi=0$ all bubbles have $R_w=0$ and this plot does not convey any information. For $\phi = 0.02$ and $0.06$
the noise arises from measuring the  curvature radius of Plateau borders. For different values of $R_{w}$,   distributions of bubble growth rates are presented in Fig. S2 of \cite{suppmat}.

\begin{figure}
       \includegraphics[angle=0,width=0.9\linewidth]{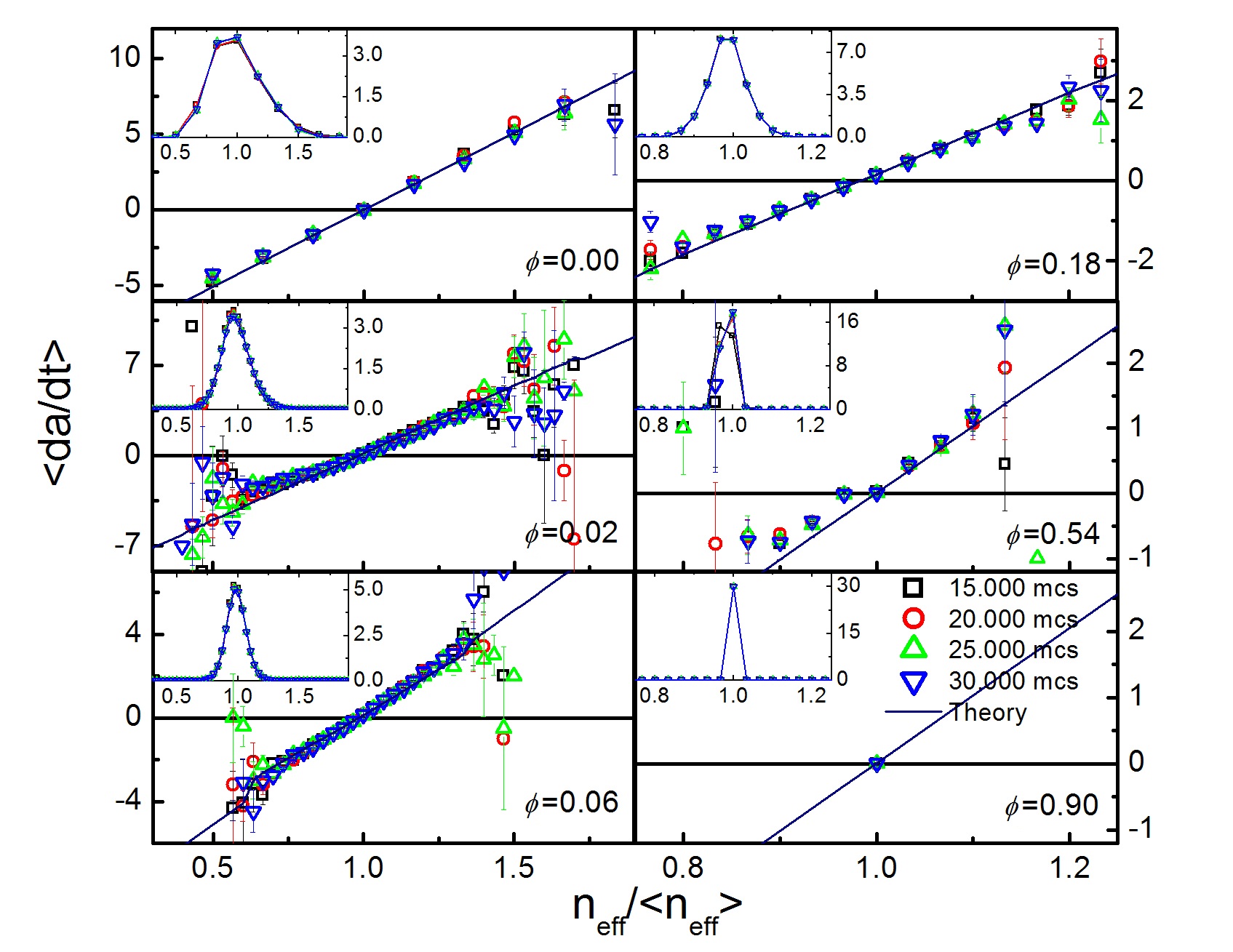}
\caption{(Color online) Gas bubble growth rate   {\it versus} $n_{eff}/\langle n_{eff}\rangle$ at different times of the self-similar growth
% scaling
 regime  for  $\phi=0$ to 0.54. Insets:  distribution function of $n_{eff}/\langle n_{eff}\rangle$.}
\label{Gr_vs_neff}
\end{figure}

Fig. \ref{Gr_vs_neff} presents the average area growth rate of gas bubbles {\it versus}  the effective number of sides, $n_{eff}$. Again, the agreement with the theoretical prediction  (eq. \ref{growth1}) is excellent.   For $\phi = 0.90$ all bubbles have  $n_{eff}=6$ and this plot does not convey any information. For different values of  $n_{eff}$, distribution functions of bubble   growth rates are plotted in Fig. S3 of  \cite{suppmat}. Plots of growth rates {\it versus} $n_{eff}/\langle n_{eff}\rangle$ (Fig. S4 of  \cite{suppmat}) and  {\it versus} $R_{w}/\langle R_{w}\rangle$ (Fig. S5 of \cite{suppmat}) discriminate the relative contributions of dry or wet interfaces to the growth.

\begin{acknowledgments}
 This work has been partially supported by Brazilian agencies CNPq, CAPES, and FAPERGS, and initiated during visits of RdA and GLT  to FG at the LSP/LIPhy, University of Grenoble.
\end{acknowledgments}

%\bibliography{wet_to_dry}

\newpage
\setcounter{figure}{0}
\onecolumngrid

\appendix
\begin{center}
{\Large\textbf{Supplementary materials}}
\end{center}

%fig1
\begin{figure}[ht]
 \includegraphics[angle=0.0,width=0.9\linewidth]{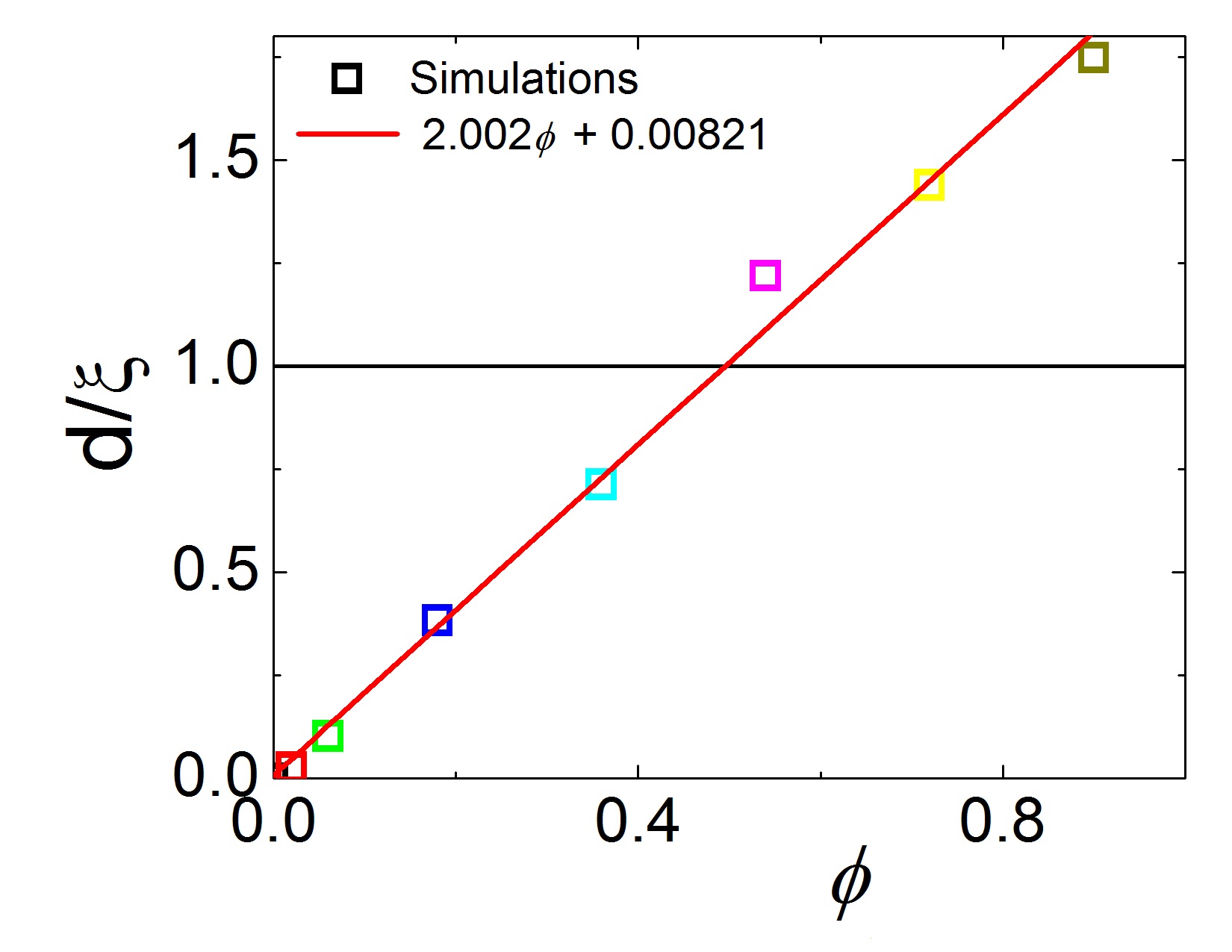}
\caption{Ratio of average distance between air bubbles and screening length $\xi$ for different liquid fractions during scaling regime. Observe that, for $\phi\geq 0.54$, $d>\xi$. The red line is a linear fit.}
\end{figure}

%fig2
\begin{figure}[ht]
  \includegraphics[angle=0.0,width=0.9\linewidth]{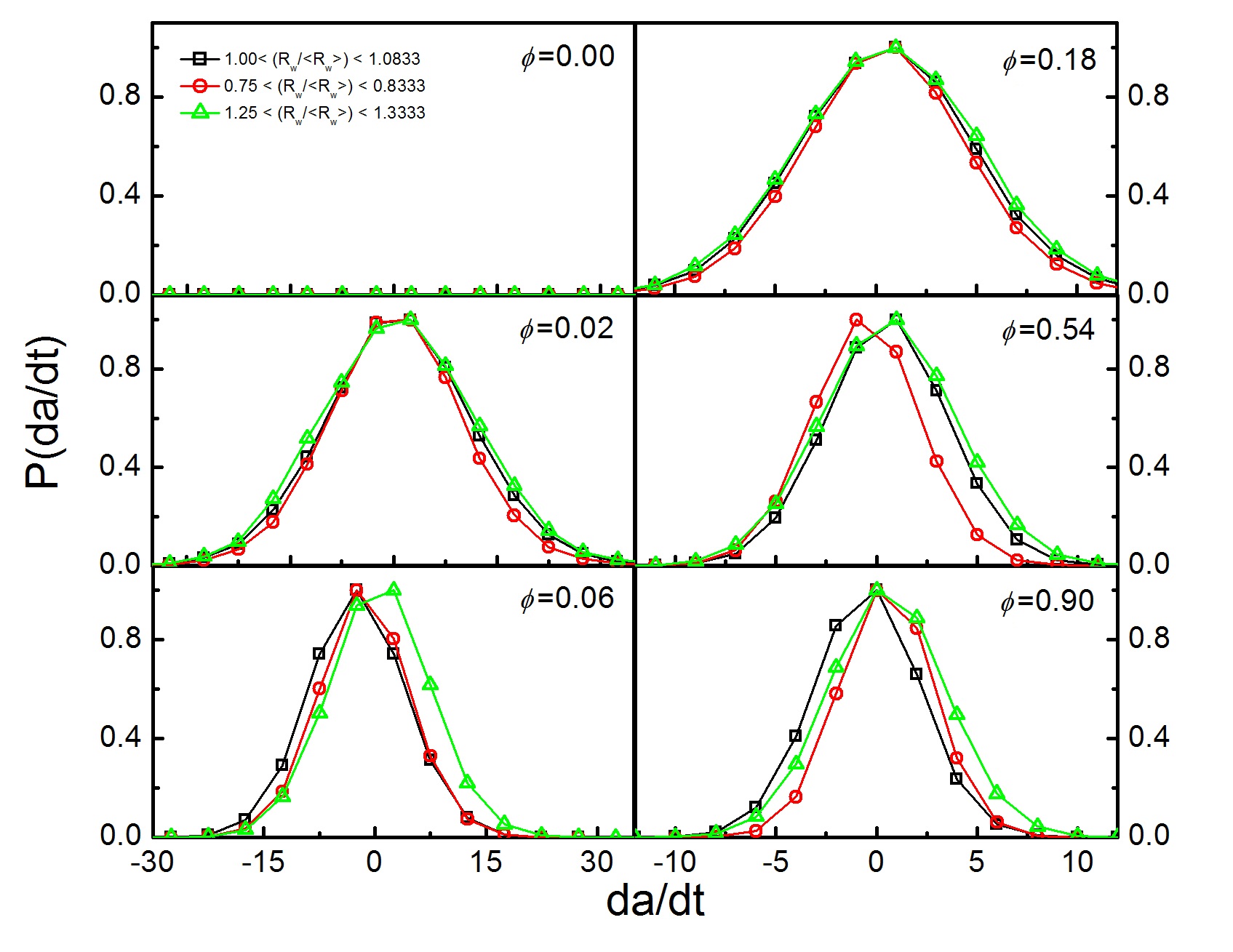}
\caption{Normalized bubble growth probability distributions, for the liquid fractions $\phi$ presented in the paper (0.0, 0.02, 0.06, 0.18, 0.54, and 0.9), in the three distinct ranges of $R_{w}/\langle R_{w}\rangle$ shown in the upper left corner. Notice that, for $\phi=0.0$,  $\langle R_{w}\rangle=0$.}
\end{figure}

%fig3
\begin{figure}[12cm]
 \includegraphics[angle=0.0,width=0.9\linewidth]{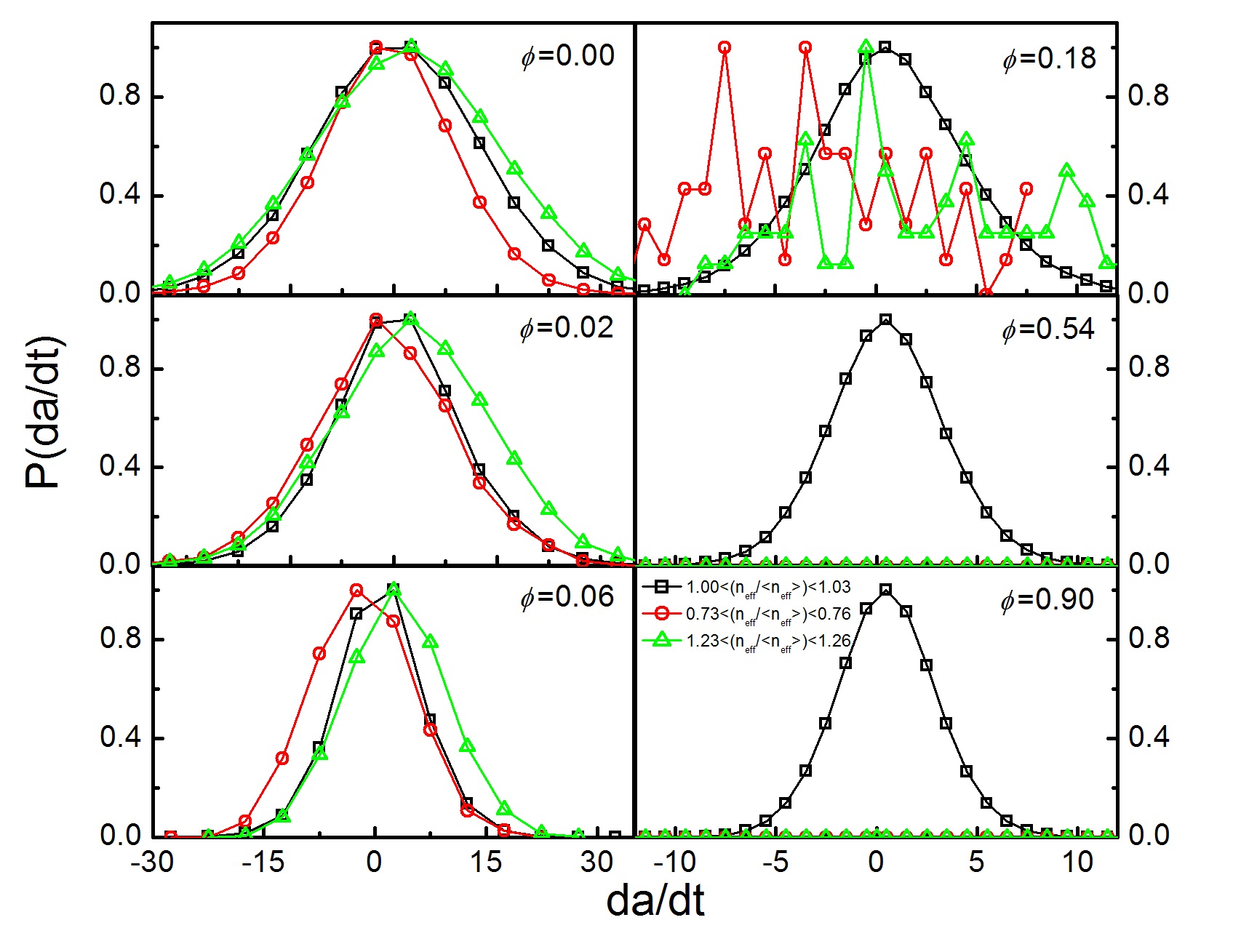}
\caption{Normalized bubble growth probability distributions, for the liquid fractions $\phi$ presented in the paper (0.0, 0.02, 0.06, 0.18, 0.54, and 0.9), in the ranges of $n_{eff}/ \langle n_{eff}\rangle$ presented in the lower right corner. Notice that, as $\phi$ increases, the distributions narrow around $n_{eff}$, so no bubbles are found far from this $n_{eff}$ value. For $\phi=0.54$ and $\phi=0.9$ there are only bubbles with $n_{eff}\simeq \langle n_{eff}\rangle$.}
\end{figure}

%fig4
\begin{figure}[12cm]
 \includegraphics[angle=0.0,width=0.9\linewidth]{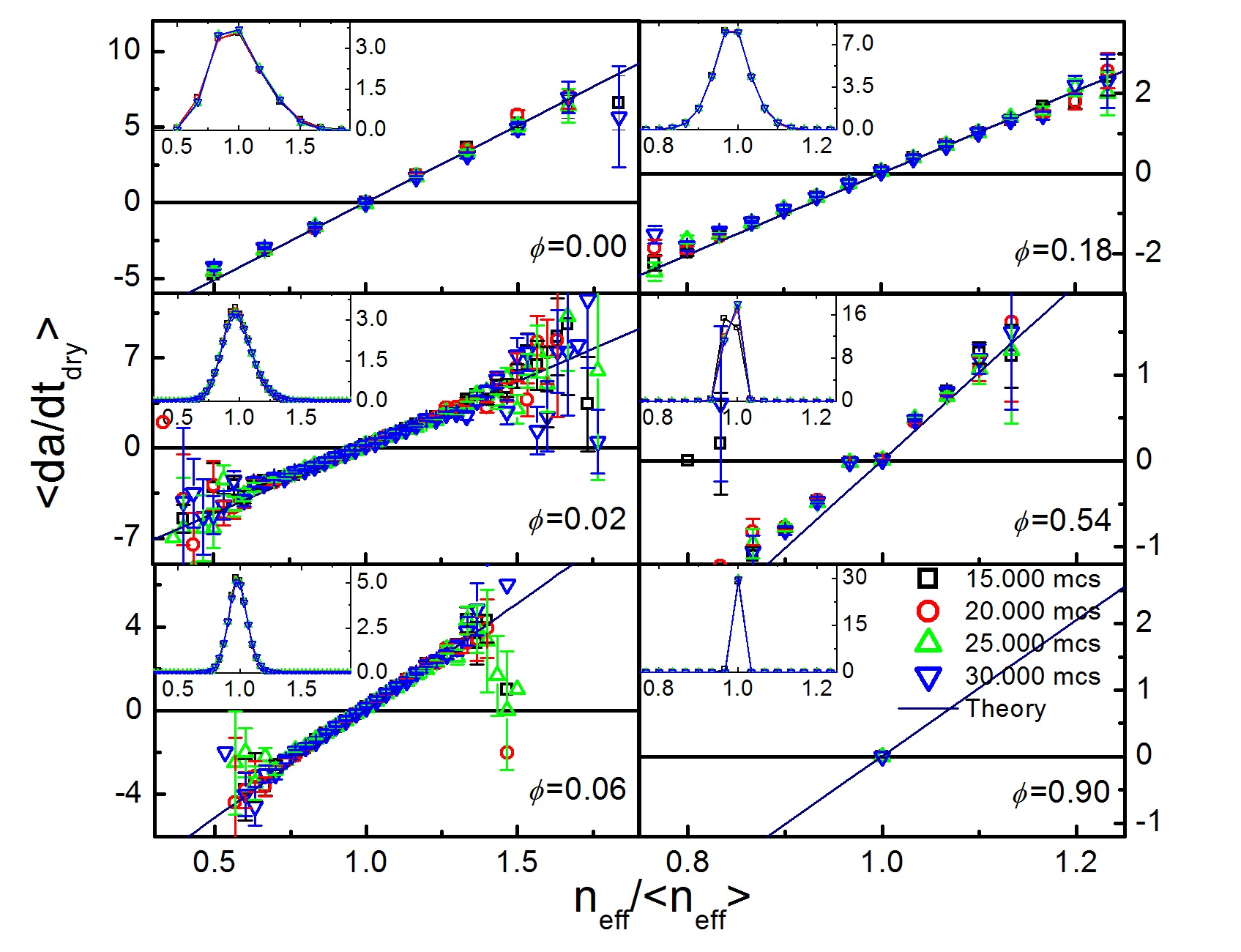}
\caption{Area growth rate of air bubbles through the dry interfaces as a function of $n_{eff}/ \langle n_{eff}\rangle$ for different liquid fractions and at different instants of the scaling regime. The insets present the probability density of $n_{eff}/ \langle n_{eff}\rangle$ at the same instants.}
\end{figure}

%fig5
\begin{figure}[12cm]
 \includegraphics[angle=0.0,width=0.9\linewidth]{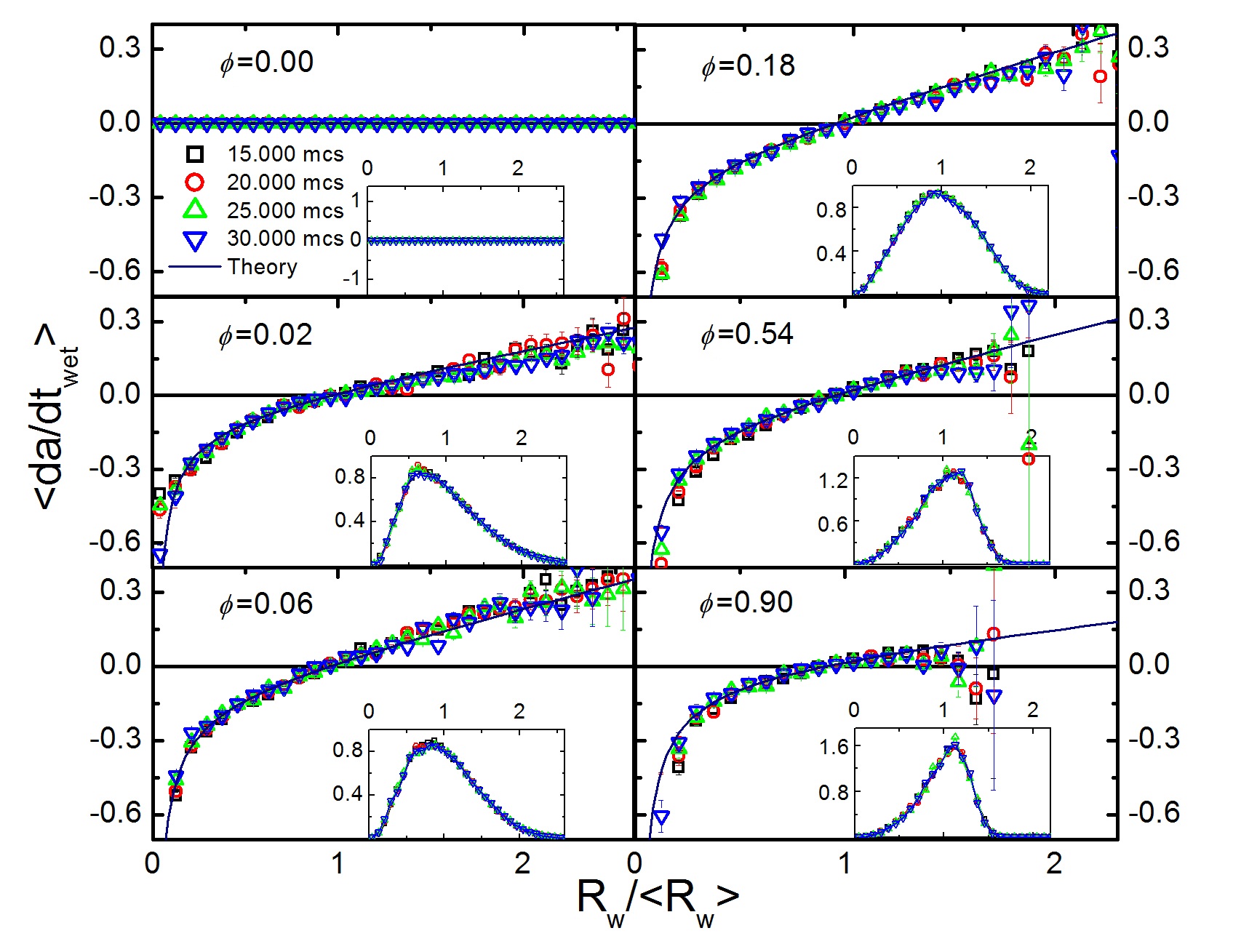}
\caption{Area growth rate of air bubbles through the wet interfaces as a function of $R_{w}/ \langle R_{w}\rangle$ for different liquid fractions and at different instants of the scaling regime. The insets present the probability density of $R_{w}/ \langle R_{w}\rangle$ at the same instants.}
\end{figure}

\end{document}